\documentclass[pra,aps,showpacs,twocolumn, a4paper,tightenlines,balancelastpage]{revtex4}
\usepackage{mathrsfs}
\usepackage{amsmath,amsfonts,amssymb,mathrsfs,bm}
\usepackage{verbatim,psfrag,times,CJK}
\usepackage{graphicx,graphics,color,epsfig}
\usepackage{flafter}
\usepackage{indentfirst}

\begin{document}

\newcommand{\ket}[1]{| #1 \rangle}
\newcommand{\bra}[1]{\langle #1 |}
\newcommand{\red}{\color[rgb]{0.8,0,0}}

\title{Two-particle dark state cooling of a nanomechanical resonator}
\author{Jia-pei Zhu$^a$}
\author{Gao-xiang Li$^{a,}$}
\email{gaox@phy.ccnu.edu.cn}
\author{Zbigniew Ficek$^b$}
\affiliation{$^{a}$Department of Physics, Huazhong Normal University, Wuhan 430079, P. R. China\\
$^{b}$The National Centre for Mathematics and Physics, KACST, P. O. Box 6086, Riyadh 11442, Saudi Arabia}

\begin{abstract}
The steady-state cooling of a nanomechanical resonator interacting with three coupled quantum dots is studied. General conditions for the cooling to the ground state with single and two-electron dark states are obtained. The results show that in the case of the interaction of the resonator with a single-electron dark state, no cooling of the resonator occurs unless the quantum dots are not identical. The steady-state cooling is possible only if the energy state of the quantum dot coupled to the drain electrode is detuned from the energy states of the dots coupled to the electron source electrode. The detuning has the effect of unequal shifting of the effective dressed states of the system that the cooling and heating processes occur at different frequencies. For the case of two electrons injected to the quantum dot system, the creation of a two-particle dark state is established to be possible with spin-antiparallel electrons. The results predict that with the two-particle dark state, an effective cooling can be achieved even with identical quantum dots subject of an asymmetry only in the charging potential energies coupling the injected electrons. It is found that similar to the case of the single-electron dark state, the asymmetries result in the cooling and heating processes to occur at different frequencies. However, an important difference between the single and two-particle dark state cases is that the cooling process occurs at significantly different frequencies. This indicates that the frequency at which the resonator could be cooled to its ground state can be changed by switching from the one-electron to the two-electron Coulomb blockade process.
\end{abstract}


\date{\today}

\maketitle

\section{Introduction}\label{sc1}

Cooling of a nanomechanical resonator (NMR) to its quantum ground state is a field of interesting theoretical and experimental research owing to their potential applications in high-precision detection of mass~\cite{s1}, mechanical displacement~\cite{s2}, and quantum information processing~\cite{s3a,s3b,s3c}. One of the conventional methods is the sideband cooling approach, based on which various cooling schemes have been put forward by coupling the resonator to a dissipative two-level system such as superconducting qubits~\cite{s4a,s4b}, nitrogen-vacancy defects in diamond~\cite{s5}, quantum dots~\cite{6,ouyang} and suspended carbon nanotubes~\cite{zippilli,mori}. Another method that could lead to an efficient cooling to the ground state involves the destructive quantum interference, i.e. the creation of a dark state that suppresses the carrier and blue sideband transitions. The method has inspired numerous cooling schemes for trapped ions such as electromagnetically induced transparency (EIT) cooling~\cite{EIT} and Stark-shift cooling~\cite{stark}. Besides these, combining the EIT cooling scheme with Stark-shift cooling scheme, a robust and fast laser cooling for trapped ions, atoms or cantilevers with closed interaction contour has been predicted~\cite{cerrillo}. It has also been demonstrated that cooling of a single NMR can be achieved via dynamical back action process or radiation pressure acting on the NMR~\cite{experiment1,experiment2}.

There have recently been a number of theoretical studies of the transport properties of a triple quantum dot (TQD) system arranged in a triangular geometry~\cite{pro1,pro2,83235319}. It has been demonstrated that by tuning of the ac field frequency, the anti-resonant behaviour in the current occurs. As the result, the stationary current is found to be sensitive to A-B effect due to spin-blockade~\cite{pro1}. In addition, the role of the dark state has been considered with two of the dots weakly connected to external leads~\cite{pro2}. The results showed that the coupling between the dot electrons and the dissipative single-phonon mode can influence the transport properties of the system~\cite{83235319}. However, it has been found that the presence of the dark state could be important not only in the study of transport properties but also for the cooling of micromechanical or nanomechanical resonators. In particular, Li {\it et al.}~\cite{epl} have recently demonstrated that an NMR can be cooled to its ground state by coupling to the TQD system operating in the strong Coulomb-blockade regime~\cite{al86}.

As we have already mentioned, the cooling schemes considered so far have been limited to the case of the strong Coulomb-blockade regime, in which at most one electron is allowed in the cooling system at a given time. However, the strong Coulomb-blockade situation may not always work since a slight change in the chemical potential of the electrodes could lead to the inclusion of two electrons into the transport window. With the possibility to inject two electrons, one can expect to see novel features in the electron transport and cooling properties of a given system. For example, P\"{o}ltl {\it et al.}~\cite{Tbrandes} have shown that the presence of two electrons in the TQD system may lead to the creation~of a two-electron dark state, which could result in breaking of the transport current and the system then exhibiting super-Poissonian behaviours~\cite{Tbrandes}.

It is the purpose of this paper to study the cooling properties of the TQD system when two electrons could be present in the dot system. Comparison is made with the single-electron cooling calculated by Li {\it et al.}~\cite{epl} that arises in the strong Coulomb-blockade regime. We include charging potential energies to describe the inclusion of two electrons. Following the prediction of P\"{o}ltl {\it et al.}~\cite{Tbrandes} that under some circumstances a two-electron dark state can be created in the dot system, we demonstrate how the presence of this dark state could lead to cooling of the NMR to its ground state. We find that in both, the single and two-electron dark state cases, the TQD system may lead to cooling of the NMR only if there is an asymmetry present in the system. In particular, in the case of single electron injected to the dot system, a cooling of the NMR occurs only when the energy state of the quantum dot couple to the drain electrode is detuned from the energy states of the dots coupled to the electron source electrode. On the other hand, with two electrons present in the dot system, cooling of the NMR can be achieved even with identical quantum dots subject of an asymmetry between the charging potential energies coupling the injected electrons.

The paper is organized as follows. In Sec.~\ref{sc2} we introduce the setup and explain the Hamiltonian of the whole system, then we derive the master equation and the rate equations for the cooling dynamics. A detailed calculation of the conditions for cooling of an NMR with single- and two-electron dark states is presented in Sec.~\ref{sc3}. We summarize our results in Sec.~\ref{sc4}. Finally, transformation matrices from the bare electron states to the dressed electron states are given in the Appendix~\ref{app}.

\section{Description of the system}\label{sc2}

\begin{figure}[h]
\begin{center}
\includegraphics[width=0.95\columnwidth]{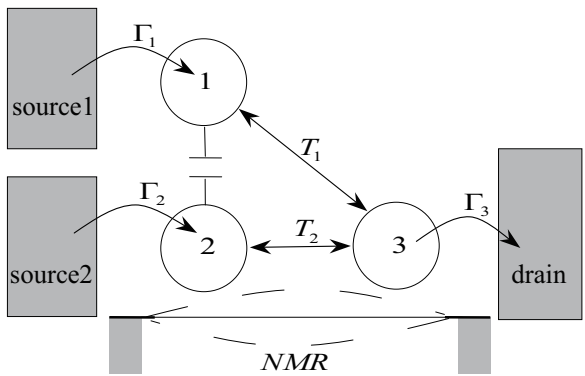}
\end{center}
\caption{Schematic diagram of the system composed of three quantum dots connected to the source and drain electrodes through tunnelling barriers. The dots $1$ and $2$ are capacitively coupled to each other without electrons tunnelling directly between them while both dots $1$ and $2$ are coupled coherently to the dot $3$ with the tunnelling amplitudes $T_{1}$ and $T_{2}$, respectively. A nanomechanical resonator (NMR) is capacitively coupled to the dots $2$ and $3$ only.}
\label{fig1}
\end{figure}

A schematic diagram of the system we consider is shown in Fig.~\ref{fig1}. Three quantum dots, each composed of a single-electron orbital state, are arranged in a triangular geometry and are weakly connected to three fermionic electrodes via tunnel barriers. Two dots, labeled $1$ and $2$, are capacitively coupled to each other without electrons tunnelling directly between them.
The dots $1$ and $2$ are coupled to the dot~$3$ through the electron tunnelling effect that occurs with amplitudes $T_{1}$ and $T_{2}$, respectively. Thus, electrons can tunnel from the dot~$1$ to $3$ and from $2$ to~$3$ with no tunnelling permitted from the dots $1$ and~$2$. The cross correlation between two capacitively coupled quantum dots in the Coulomb blockade regime has recently been measured~\cite{capaci}. We assume that only two quantum dots, $2$ and $3$ are directly coupled to the NMR.

\subsection{Hamiltonian of the system}

We can divide the Hamiltonian of the entire system into four parts:~(1) Three coupled quantum dots, (2) three normal metal electrodes coupled to the quantum dots, (3) a mechanical resonator interacting with two of the three quantum dots, and (4) a reservoir or thermal bath surrounding the mechanical resonator.

The dynamics of the three coupled quantum dots containing maximum two electrons can be determined by the Hamiltonian
\begin{align}
\hat{H}_{d} &= \sum_{i=1}^{3}\sum_{\sigma}E_{i}\hat{n}_{i\sigma}+\sum_{\sigma}\!\left(T_1\hat{d}_{1\sigma}^{\dagger}\hat{d}_{3\sigma}+T_2\hat{d}_{2\sigma}^{\dagger}\hat{d}_{3\sigma}+ {\rm H.c.}\!\right)\nonumber\\
&+\sum_{i=1}^{3}U_{ii}\hat{n}_{i\uparrow}\hat{n}_{i\downarrow}+\sum_{i<j=1}^{3}\sum_{\sigma,\sigma'}U_{ij}\hat{n}_{i\sigma}\hat{n}_{j\sigma'} ,\label{Hd}
\end{align}
where $\hat{n}_{i\sigma}=\hat{d}_{i\sigma}^{\dagger}\hat{d}_{i\sigma}$ is a number operator describing the number of spin-up $(\sigma = \uparrow)$ and spin-down $(\sigma =\downarrow)$ electrons in the $i$th quantum dot. The operators $\hat{d}_{i\sigma}$ and $\hat{d}_{i\sigma}^{\dagger}$ are respectively the annihilation  and creation operators of an electron with spin~$\sigma$ in the $i$th quantum dot, and $E_{i}$ is the energy of the single-electron state in the dot $i$. We assume that the energy states of the electron in each quantum dot are spin-independent, so both spin-up and spin-down electrons can equally occupy the energy states of the quantum dots. The coupling between the quantum dots is arranged through a tunnelling process such that the tunnelling of electrons between the dots $1$ and $3$ occurs with an amplitude $T_{1}$, whereas between the dots $2$ and~$3$ occurs with an amplitude $T_{2}$. There is no direct tunnelling allowed between the dots $1$ and $2$. In this case, the system may be considered as a Lambda-type system with two ground states $1$ and $2$ and the upper state $3$.

A possibility of simultaneous injection of two electrons into the dots is determined in the Hamiltonian (\ref{Hd}) by charging potential energies. There are two distinct types of the charging potential energies $U_{ii}$ and $U_{ij}\, (i\neq j)$, corresponding to the charging energy of two electrons injected to the same $i$th dot and to different dots, respectively. Notice that the charging potential energies $U_{ij}$ are independent of the polarization of the spins.

The electrons can be injected to the quantum dots from normal metal electrodes by coupling them through a tunnelling process. Suppose that the electrodes coupled to the dots $1$ and~$2$ appear as a source (donors) of the electrons while the electrode coupled to the dot $3$ appears as an absorber (drain) of the electrons. In this case, the Hamiltonian of the electrodes coupled to the quantum dots can be written as
\begin{align}
\hat{H}_{e} =\sum\limits_{j=1}^{3}\sum\limits_{k,\sigma}\left[\epsilon_{jk} \hat{c}^{\dag}_{jk\sigma}\hat{c}_{jk\sigma}
+ \left(V_{jk} \hat{c}^{\dag}_{jk\sigma}\hat{d}_{j\sigma} + {\rm H.c.}\right)\right] ,\label{He}
\end{align}
where the first term is the energy of the electrons with $\epsilon_{jk}$ representing the energy of a single electron of wave number~$k$ in the electrode $j$. The operators $\hat{c}_{ik\sigma}$ and $\hat{c}_{ik\sigma}^{\dagger}$ are respectively the annihilation and creation operators of an electron with spin~$\sigma$ and wave number~$k$ in the electrode $j$. The second term in Eq.~(\ref{He}) is the interaction between the electrodes and the quantum dots with $V_{jk}$ denoting the coupling strength between an electron of the wave number $k$ in the $j$th electrode and the corresponding quantum~dot. The coupling is arranged through a tunnelling of the electrons between the electrodes and the quantum dots, and is independent of $\sigma$, so that both spin-up and spin-down electrons are coupled equally to the quantum dots.

The capacitor adjacent to the dots is formed by the NMR and a static plate with a gate voltage applied to it. The effective capacitance of this capacitor could be adjusted by the displacement of the NMR from its equilibrium position~\cite{ouyang,epl}. In this way, the displacement of the NMR couples to the number of electrons in the dots. We assume that only two of them, the dots $2$ and $3$ participate in the interaction with the NMR. In this case, the Hamiltonian of the NMR interacting with the quantum dots is of the form~\cite{e-p,mori}
\begin{align}
\hat{H}_{m} = \omega_m \hat{a}^{\dagger}\hat{a} + \sum_\sigma \left(\hat{a}+\hat{a}^{\dagger}\right)\left(\alpha_{1}\hat{n}_{2\sigma}+\alpha_{2}\hat{n}_{3\sigma}\right) ,\label{epcoupling}
\end{align}
where $\hat{a}^{\dagger}$ and $\hat{a}$ are operators representing the creation and annihilation of a phonon excitation of frequency $\omega_m$ in the~NMR, and $\alpha_{i}\, (i=1,2)$ is the electromechanical coupling constant between the $i$th quantum dot and the phonon mode. Note that the interaction of the NMR with the dots is in a form of the nonlinear parametric interaction, known in optical systems as a radiation pressure type interaction~\cite{genes}.

In addition to the interaction with the quantum dots, the phonon mode of the NMR is also allowed to interact with a heat bath or thermal reservoir which results in the presence of thermal phonons in the NMR mode. The Hamiltonian for this case is simply written as
\begin{eqnarray}
\hat{H}_{b} = \sum\limits_q\omega_q \hat{a}^{\dagger}_q \hat{a}_q+\sum\limits_q g_q\left(\hat{a}^{\dagger}_q \hat{a}+\hat{a}_q\hat{a}^{\dagger}\right) ,\label{Hres}
\end{eqnarray}
where $\hat{a}^{\dagger}_q$ and $\hat{a}_{q}$ are respectively the creation and annihilation operators of the $q$th mode of frequency $\omega_{q}$ of the multi-mode reservoir, and $g_q$ is the coupling strength of the mode $q$ to the~NMR phonon mode $\hat{a}$.

We should point out here that in the Hamiltonians (\ref{Hd})-(\ref{Hres}) the coupling parts retain only the terms which play a dominant role in the rotating-wave approximation. Antiresonant terms which would make much smaller contributions have been omitted.

\subsection{One and two electron dynamics}

The possibility of a simultaneous injection of two electrons to the quantum dots system can be directly addressed by writing the creation operators $\hat{d}_{i\sigma}^{\dagger}$ in the two-electron basis as
\begin{align}
\hat{d}_{1\uparrow}^{\dagger}&= |1_\uparrow 0\rangle\langle00| + |1_\uparrow2_\uparrow\rangle\langle2_\uparrow 0| + |1_\uparrow3_\uparrow\rangle\langle3_\uparrow 0|\!+\!|1_\uparrow1_\downarrow\rangle\langle1_\downarrow 0| \nonumber\\
&+|1_\uparrow2_\downarrow\rangle\langle2_\downarrow 0| + |1_\uparrow3_\downarrow\rangle\langle3_\downarrow0| ,\nonumber\\
\hat{d}_{2\uparrow}^{\dagger}&= |2_\uparrow0\rangle\langle00| - |1_\uparrow2_\uparrow\rangle\langle1_\uparrow0| + |2_\uparrow3_\uparrow\rangle\langle3_\uparrow0|\!+\!|2_\uparrow1_\downarrow\rangle\langle1_\downarrow0| \nonumber\\
&+ |2_\uparrow2_\downarrow\rangle\langle2_\downarrow0| + |2_\uparrow3_\downarrow\rangle\langle3_\downarrow0| ,\nonumber\\
\hat{d}_{3\uparrow}^{\dagger}&= |3_\uparrow0\rangle\langle00| - |1_\uparrow3_\uparrow\rangle\langle1_\uparrow0| - |2_\uparrow3_\uparrow\rangle\langle2_\uparrow0|\!+\!|3_\uparrow1_\downarrow\rangle\langle1_\downarrow0| \nonumber\\
&+ |3_\uparrow2_\downarrow\rangle\langle2_\downarrow0| + |3_\uparrow3_\downarrow\rangle\langle3_\downarrow0| ,\nonumber\\
\hat{d}_{1\downarrow}^{\dagger}&= |1_\downarrow0\rangle\langle00| - |1_\uparrow1_\downarrow\rangle\langle1_\uparrow0| - |2_\uparrow1_\downarrow\rangle\langle2_\uparrow0|\!-\!|3_\uparrow1_\downarrow\rangle\langle3_\uparrow0|\nonumber\\
&+ |1_\downarrow2_\downarrow\rangle\langle2_\downarrow0| + |1_\downarrow3_\downarrow\rangle\langle3_\downarrow0|,\nonumber\\
\hat{d}_{2\downarrow}^{\dagger}&= |2_\downarrow0\rangle\langle00| - |1_\uparrow2_\downarrow\rangle\langle1_\uparrow0| - |2_\uparrow2_\downarrow\rangle\langle2_\uparrow0|\!-\!|3_\uparrow2_\downarrow\rangle\langle3_\uparrow0|\nonumber\\
&- |1_\downarrow2_\downarrow\rangle\langle1_\downarrow0| + |2_\downarrow3_\downarrow\rangle\langle3_\downarrow0|,\nonumber\\
\hat{d}_{3\downarrow}^{\dagger}&= |3_\downarrow0\rangle\langle00| - |1_\uparrow3_\downarrow\rangle\langle1_\uparrow0| - |2_\uparrow3_\downarrow\rangle\langle2_\uparrow0|\!-\!|3_\uparrow3_\downarrow\rangle\langle3_\uparrow0| \nonumber\\
&- |1_\downarrow3_\downarrow\rangle\langle1_\downarrow0| - |2_\downarrow3_\downarrow\rangle\langle2_\downarrow0| ,\label{operators}
\end{align}
where $\ket{i_\sigma j_{\sigma^{\prime}}}\equiv \ket{i_\sigma}\otimes\ket{j_{\sigma^{\prime}}}$ is a product state of single electron states in which dots $i$ and~$j$ are occupied by an electron of spin $\sigma,\sigma^{\prime}=\uparrow,\downarrow$. The state $\ket 0$ corresponds to the case of no electron occupying any of the quantum dots. The minus signs in the expressions arise from the anti-symmetric property of the fermion system with respect to the exchange of two electrons.

We may then decompose the Hamiltonian $\hat{H}_{d}$ into one- and two-electron parts
\begin{eqnarray}
\hat{H}_{d} = \sum_{\sigma}\hat{H}_{\sigma} +\sum_{\sigma}\hat{H}_{\sigma\sigma} +\sum_{\sigma\neq\sigma^{\prime}}\hat{H}_{\sigma\sigma^{\prime}} ,\label{Hd2}
\end{eqnarray}
where $\hat{H}_{\sigma}$ represents the energy of the quantum dot system containing only a single electron of spin $\sigma$, the Hamiltonian~$\hat{H}_{\sigma\sigma}$ represents the case when two parallel-spin electrons are injected to the quantum dots, and $\hat{H}_{\sigma\sigma^{\prime}}\, (\sigma\neq \sigma^{\prime})$ represents the case of two opposite-spin electrons present in the quantum dot system.

The single- and two-electron Hamiltonians can be conveniently written in matrix forms. The space of the quantum dot system containing a single electron is spanned by three state vectors, ${|1_{\sigma}\rangle,|2_{\sigma}\rangle,|3_{\sigma}\rangle}$, the space of the system containing two parallel-spin electrons is spanned by three state vectors, ${|1_{\sigma}2_{\sigma}\rangle,|1_{\sigma}3_{\sigma}\rangle,|2_{\sigma}3_{\sigma}\rangle}$, and the space of the system containing two opposite-spin electrons is spanned by nine state vectors, $|1_{\uparrow}1_{\downarrow}\rangle,$ $|1_{\uparrow}2_{\downarrow}\rangle,$ $|2_{\uparrow}1_{\downarrow}\rangle,$ $|2_{\uparrow}2_{\downarrow}\rangle,$ $|3_{\uparrow}3_{\downarrow}\rangle$, $|1_{\uparrow}3_{\downarrow}\rangle,$ $|2_{\uparrow}3_{\downarrow}\rangle,$ $|3_{\uparrow}1_{\downarrow}\rangle,$ $|3_{\uparrow}2_{\downarrow}\rangle$. It is then straightforward to show that the Hamiltonian $\hat{H}_{\sigma}$ written in the single-spin basis is of the form
\begin{align}
\hat{H}_{\sigma}=\left(
\begin{array}{ccc}
    0 & 0 & T_{1} \\
    0 & \Delta_{1} & T_{2} \\
    T_{1} & T_{2} & \Delta_{2} \\
  \end{array}
\right) + E_{1}I ,\label{hsi}
\end{align}
where $\Delta_{1}=E_{2}-E_{1}$ and $\Delta_{2}=E_{3}-E_{1}$ are energy differences between the dot $1$ and the dots $2$ and $3$, respectively, and a single energy factor $E_{1}I$, in which $I$ is the unit matrix, has been pulled out for the convenience of diagonalization of the Hamiltonian and interpretation. This term can be referred as a reference energy about which the energies of the states are centered.

The Hamiltonian $\hat{H}_{\sigma\sigma}$ written in the basis of two parallel-spin electrons is of the form
\begin{align}
\hat{H}_{\sigma\sigma}\!=\!\left(
\begin{array}{ccc}
    0 & T_{1} & - T_{2} \\
    T_{1} & \Delta_{1}\!+\!\delta_{1} & 0 \\
    - T_{2} & 0 & \Delta_{2}\!+\!\delta_{2} \\
  \end{array}
\right)\!+\!(E_{1}\!+\!E_{2}\!+\!U_{12})I ,\label{hspa}
\end{align}
where $\delta_{1}=U_{13}-U_{12}$ and $\delta_{2}=U_{23}-U_{12}$.

Finally, the Hamiltonian $\hat{H}_{\sigma\sigma^{\prime}}$ written in the basis of two opposite-spin electrons is represented by the following $9\times 9$ matrix
\begin{align}
\hat{H}_{\sigma\sigma^{\prime}}&=\left(
  \begin{array}{ccccccccc}
    \delta_{11}& 0 & 0 & 0 & 0 & T & 0 &T & 0 \\
    0 & 0 & 0 & 0 & 0 & T & 0 & 0 & T \\
    0 & 0 &0 & 0 & 0 & 0 & T & T & 0 \\
    0 & 0 & 0 & \delta_{22} & 0 & 0 & T & 0 & T \\
    0 & 0 & 0 & 0 & \delta_{33} & T& T & T & T \\
    T & T & 0 & 0 & T & \delta_{13} & 0 & 0 & 0 \\
    0 & 0 & T & T & T & 0 & \delta_{23} & 0 & 0 \\
    T & 0 & T & 0 & T & 0 & 0 & \delta_{13} & 0 \\
    0 & T & 0 & T & T & 0 & 0 & 0 & \delta_{23} \\
  \end{array}
\right) \nonumber\\
&+(E_{1}+E_{2}+U_{12})I ,\label{oppo}
\end{align}
where
\begin{align}
\delta_{11} &= -\Delta_{1}+U_{11}-U_{12} ,\quad \delta_{22} =\Delta_{1}+U_{22}-U_{12} ,\nonumber\\
\delta_{33} &= \Delta_{2}+\Delta_{3}+U_{33}-U_{12} ,\quad \delta_{13} =\Delta_{2}+\delta_{1} ,\nonumber\\
\delta_{23} &= \Delta_{3} +\delta_{2} ,
\end{align}
with $\Delta_{3}=E_{3}-E_{2}$.

\subsection{Master equation}

We now suppose that the quantum dots and the NMR are weakly coupled to the electrodes and to the thermal reservoir, respectively. This "reduced" quantum system, i.e. the quantum dots plus the NMR can be described by a reduced density operator $\rho$, which is obtained by tracing the density matrix of the entire system $\rho_{T}$ over the space of the $\hat{c}_{jk\sigma}$ and~$\hat{a}_{q}$ modes~\cite{Tbrandes,Platero1,Platero2,Liao}
\begin{align}
\rho ={\rm Tr}_{(e,r)}\rho_{T} .
\end{align}
We shall derive the master equation for this reduced density operator by assuming that the reservoir modes and the electrons modes in the electrodes are $\delta$-correlated, i.e. they appear as white (frequency independent) noises to the NMR and to the quantum dots, respectively.
If we limit the interaction of the reduced system to the second-order in the coupling constants~$V_{jk}$ and $g_{q}$, we then find that under the Born-Markov and the rotating-wave approximations, the the reduced density matrix $\rho$ satisfies the master equation~\cite{lindblad}
\begin{eqnarray}
\frac{d}{dt}\rho = -i[\hat{H}_{d} +\hat{H}_{m},\rho]+\mathcal{L}_p\rho+\mathcal {L}_e\rho ,\label{master}
\end{eqnarray}
in which
\begin{align}
\mathcal {L}_p\rho &= \frac{1}{2}(\bar{n}_p+1)\gamma_p\left(2\hat{a}\rho \hat{a}^{\dagger}-\hat{a}^{\dagger}\hat{a}\rho-\rho \hat{a}^{\dagger}\hat{a}\right)\nonumber\\
&+\frac{1}{2}\bar{n}_p\gamma_p\left(2\hat{a}^{\dagger}\rho \hat{a}- \hat{a}\hat{a}^{\dagger}\rho- \rho \hat{a}\hat{a}^{\dagger}\right) ,\nonumber\\
\mathcal{L}_e\rho &= \frac{1}{2}\sum\limits_{i=1}^{2}\sum_{\sigma}\Gamma_{i\sigma}\left(2\hat{d}^{\dagger}_{i\sigma}\rho \hat{d}_{i\sigma}- \hat{d}_{i\sigma}\hat{d}^{\dagger}_{i\sigma}\rho - \rho\hat{d}_{i\sigma}\hat{d}^{\dagger}_{i\sigma}\right) \nonumber\\
&+\frac{1}{2}\sum_{\sigma}\Gamma_{3\sigma}\!\left(2\hat{d}_{3\sigma}\rho \hat{d}^{\dagger}_{3\sigma}- \hat{d}^{\dagger}_{3\sigma}\hat{d}_{3\sigma}\rho-\rho \hat{d}^{\dagger}_{3\sigma}\hat{d}_{3\sigma}\!\right) ,\label{equ13}
\end{align}
where $\mathcal{L}_p\rho$ is an operator representing processes that lead to thermalization of the NMR phonon mode with the rate $\gamma_{p}$, and $\bar{n}_p=[\exp(\omega_m/k_{B}T_{p})-1]^{-1}$ is the number of thermal phonons of frequency $\omega_m$ at temperature $T_{p}$. The operator~$\mathcal{L}_e\rho$ represents the one-way tunnelling process of injection of electrons to the quantum dots $1$ and $2$ from the source electrodes with chemical potential $\mu_{1,2}$, and the damping of the dot $3$ by the one-way electron tunnelling out to the drain electrode with chemical potential $\mu_3$. The electrons tunnel into the dots~$1$ and~$2$ with rates $\Gamma_{1\sigma}$ and $\Gamma_{2\sigma}$, respectively, whereas they tunnel away from the dot $3$ with the rate $\Gamma_{3\sigma}$, as illustrated in Fig.~\ref{fig1}.

The difference between the source and drain electrodes in the master equation~(\ref{master}) results from the difference between the corresponding chemical potentials, although the Hamiltonian described in Eq.~(\ref{He}) is identical for all electrodes. In the derivation, we have assumed that the chemical potentials $\mu_{1,2}$ and $\mu_3$ obey the relations, $\mu_{1,2} > E_{1,2}$ and $\mu_3 < E_3$, which are required for the tunnelling processes to occur in one direction. For these chemical potentials in the equilibrium reservoirs, the Fermi distributions in the source and drain electrodes can well be approximated by $f_i(E_i)=1\, (i=1,2)$, and $f_3(E_3)=0$. Thus, in the infinite-bias limit of~$\mu_{1,2}\rightarrow\infty$ and $\mu_3\rightarrow -\infty$, the three tunnelling rates have the form
\begin{align}
\Gamma_{i\sigma} = 2\pi\sum\limits_{k}|V_{ik}|^2\delta(E_i-\epsilon_{ik}) ,
\end{align}
which is assumed to be independent of the energy $E_i$ and spin~$\sigma$.

\subsection{Rate equation for the average phonon number}

Having derived the master equation for the NMR coupled to the quantum dots system, we now turn our attention to the derivation of a rate equation for the average phonon number in the resonator modes. In doing this, we will assume that the coupling between the quantum dots and the resonator is weak, $\alpha_{1,2}\ll\omega_{m},T_{1,2}$. In such a case, we may use the second-order perturbation theory to eliminate the degrees of the freedom of the quantum dot system. A similar procedure has been used in the framework of single-electron dark states~\cite{ouyang,zippilli,cerrillo,JPB}.

We may extract from the Hamiltonian (\ref{epcoupling}) the part describing the interaction between the dots $2$ and $3$ and the NMR
\begin{align}
\hat{H}_{\alpha} =  \alpha \sum_\sigma \left(\hat{a}+\hat{a}^{\dagger}\right)\left(\hat{n}_{2\sigma}+\hat{n}_{3\sigma}\right) ,\label{hal}
\end{align}
where, for simplicity, we put $\alpha_{1}=\alpha_{2}\equiv \alpha$, and write the master equation~(\ref{master}) in the form
\begin{equation}
\frac{d}{dt}\rho = \mathcal{L}_0\rho +\mathcal{L}_{\alpha}\rho ,\label{rate}
\end{equation}
in which the Liouvillian
\begin{align}
\mathcal{L}_0\rho = -i[\hat{H}_{d},\rho] -i\omega_{m}[\hat{a}^{\dagger}\hat{a},\rho] +\mathcal{L}_e\rho +\mathcal{L}_{p}\rho
\end{align}
represents the dynamics of the quantum dots and the phonon mode in the absence of the electron-phonon coupling, and
\begin{align}
\mathcal{L}_{\alpha}\rho=-i[\hat{H}_{\alpha},\rho] \label{inte}
\end{align}
is the interaction between the quantum dots and the phonon mode. In the limit of $\alpha\ll \omega_{m}$, the interaction (\ref{inte}) can be treated as a weak perturber to the system.

We now follow the procedure of Cirac {\it et al.}~\cite{cb92} of using the second-order perturbation theory in respect to the coupling strength~$\alpha$, to obtain from the master equation (\ref{rate}) a set of rate equations for the occupation probabilities $p_{n}=\bra n \rho\ket n$ of the NMR phonon state with~$n$ excitations. Straightforward but lengthly calculations yield the following equation
\begin{align}
\frac{d}{dt}p_n &= \left[(\bar{n}_p+1)\gamma_p +A_{-}\right][(n+1)p_{n+1}-np_{n}] \nonumber\\
&+(\bar{n}_p\gamma_p +A_{+})[np_{n-1}-(n+1)p_{n}] ,\label{e7}
\end{align}
in which the transition rates $A_{\pm}$ are given by
\begin{align}
A_{\pm} &= 2\alpha^2{\rm Re}S(\pm\omega_m) \nonumber\\
&= 2\alpha^2{\rm Re}\int\limits_0^\infty d\tau\langle\hat{\mathcal {O}}(\tau)\hat{\mathcal {O}}(0)\rangle_{s} {\rm e}^{\mp i\omega_m\tau} ,\label{correlation}
\end{align}
where $\langle\hat{\mathcal {O}}(\tau)\hat{\mathcal {O}}(0)\rangle_{s}$ is the correlation function for the quantum dots operators $\hat{\mathcal {O}}(\tau)=\sum_{\sigma}[\hat{n}_{2\sigma}(\tau)+\hat{n}_{3\sigma}(\tau)]$, evaluated in the absence of the coupling $\alpha$. The two-time correlation function is evaluated using the quantum regression theorem~\cite{lax}, which states that the two-time correlation function $\langle\hat{\mathcal {O}}(\tau)\hat{\mathcal {O}}(0)\rangle_{s}$ satisfies the same equations of motion for $\tau >0$ as the corresponding one-time averages $\langle\hat{\mathcal {O}}(\tau)\rangle$.  In turn, the one-time averages can be expressed in terms of the matrix elements of the reduced density operator of the the quantum dots~$\rho_{d} ={\rm Tr}_{r}\rho$, where the trace is taken over the space of the NMR mode.

To examine the occurrence of cooling we look at the average number of phonons in the steady-state of the NMR mode, $\langle n\rangle =\sum_{n}np_n$. Using Eq.~(\ref{e7}), we obtain the following simple equation of motion
\begin{eqnarray}
\frac{d}{dt}\langle n\rangle = -\left(\gamma_p+A_{-}-A_{+}\right)\langle n\rangle+\gamma_p\bar{n}_p+A_{+} .\label{dn}
\end{eqnarray}
In order to calculate $\langle n\rangle$ we need to determine the rates $A_{\pm}$. It is seen from Eq.~(\ref{dn})
that $A_{+}$ and $A_{-}$ are heating and cooling rates, respectively. They represent processes which can increase (heating) and decrease (cooling) of an excitation of the NMR mode when at most two electrons are simultaneously injected to the quantum dots. Under the condition of $A_->A_+$, the steady-state average number of phonons is of the form
\begin{align}
\langle n\rangle_{s} =\frac{\gamma_p\bar{n}_p+A_{+}}{\gamma_p+A_{-}-A_{+}} .\label{nav}
\end{align}
It is clear by examination of Eq.~(\ref{nav}) that a significant cooling of the NMR can be achieved when $A_{-}\gg \gamma_{p},A_{+}$.

In the following section, we shall use Eq.~(\ref{nav}) to study the conditions for cooling of the NMR to its ground state with the help of the two-particle dark state. As we shall see, electrons trapped in the two-particle dark state can be used as the carriers to transfer the phonon energy from the NMR mode to the drain, thereby cooling the NMR to its ground state.

\section{Dark-state cooling of the NMR}~\label{sc3}

Now we demonstrate how to apply the two-particle dark to cool the NMR to its ground state. The task is to determine the mean phonon number in the steady-state which expresses the information about temperature of the oscillating resonator. To see the advantages of using the two-particle dark state, we first briefly consider the cooling mechanism with a single-particle dark state~\cite{epl,74125315,76245319,83235319}.

\subsection{Cooling with the single-electron dark state}

In the strong Coulomb-blockade regime, at most one electron can tunnel into the quantum dot system. With only one electron present, the quantum-dot system is described by the Hamiltonian $H_{\sigma}$ alone, Eq.~(\ref{hsi}). We may diagonalize the $3\times 3$ matrix appearing in Eq.~(\ref{hsi}) to find the effective (dressed) states of the electron. On calculating the eigenvalues and eigenvectors of the matrix one can find that all of the eigenvalues are different from zero and the corresponding eigenvectors all involve the state $\ket{3_{\sigma}}$. This indicates that in general no electron trapping (dark) state is created because the electron can escape from the dot~$3$ to the drain electrode. However, it is easily verified that in the limit of $\Delta_{1}=0$, the eigenvalues of the matrix are
\begin{align}
\lambda_{1}=0 ,\quad \lambda_{\pm} = \frac{1}{2}\Delta_{2}\pm \sqrt{\left(T_{1}^{2}+T_{2}^{2}\right) +\frac{1}{4}\Delta_{2}^{2}} ,\label{dei}
\end{align}
and the corresponding eigenvectors are of the form
\begin{align}
|\varphi_{1\sigma}\rangle &= \cos\eta\ket{1_{\sigma}} - \sin\eta\ket{2_{\sigma}} ,\nonumber\\
|\varphi_{+\sigma}\rangle &= \cos\theta|3_{\sigma}\rangle + \sin\theta(\sin\eta\ket{1_{\sigma}} +\cos\eta\ket{2_{\sigma}}) ,\nonumber\\
|\varphi_{-\sigma}\rangle &= \sin\theta|3_{\sigma}\rangle - \cos\theta(\sin\eta |1_{\sigma}\rangle+\cos\eta |2_{\sigma}\rangle) ,\label{single}
\end{align}
where $\tan\eta =T_{1}/T_{2}$, and
\begin{align}
\tan\theta = \left(\frac{\Omega -\Delta_{2}}{\Omega +\Delta_{2}}\right)^{\frac{1}{2}} ,
\end{align}
with $\Omega = \sqrt{\Delta_{2}^{2} +4(T^{2}_{1}+T_{2}^{2})}$.

We see that one of the eigenvalues $(\lambda_{1})$ is zero and the eigenvector corresponding to $\lambda_{1}$ does not involve the state~$\ket{3_{\sigma}}$. This shows that by choosing two pumped dots of the same energies, $E_{2}=E_{1}$, we may achieve a dark state that is decoupled from the dot~$3$. This statement remains true even if $\Delta_{2}=0$. Note that the creation of the dark state is independent of the ratio $T_{1}/T_{2}$.

It is straightforward to see from Eq.~(\ref{dei}) that $\lambda_{+}>\lambda_{1}>\lambda_{-}$. Thus, the state $\ket{\varphi_{+\sigma}}$ is shifted upwards from the dark state energy by the amount $\lambda_{+}$, whereas the state $\ket{\varphi_{-\sigma}}$ is shifted downwards from $\ket{\varphi_{1\sigma}}$ by the amount $-\lambda_{-}$. Hence, the electron will experience gain of its energy when making a transition from the dark state $\ket{\varphi_{1\sigma}}$ to the state~$\ket{\varphi_{+\sigma}}$ and absorption of the energy when making a transition from the dark state to the state $\ket{\varphi_{-\sigma}}$. Thus, cooling of the NMR is possible only when the electron interacting with the NMR makes transition to the upper state. When $\Delta_{2}=0$, i.e. all three dots are identical, the states $\ket{\varphi_{+\sigma}}$ and $\ket{\varphi_{-\sigma}}$ are then symmetrically located about the dark state $\ket{\varphi_{1\sigma}}$. In this case, the transition frequencies $\ket{\varphi_{1\sigma}}\rightarrow\ket{\varphi_{+\sigma}}$ and $\ket{\varphi_{1\sigma}}\rightarrow\ket{\varphi_{-\sigma}}$ overlap. Consequently, the cooling and heating frequencies overlap resulting in no effective reduction of the average number of phonons. Hence, the one-electron cooling can be observed only when $\Delta_{2}\neq 0$, i.e. when the states $\ket{\varphi_{+\sigma}}$ and $\ket{\varphi_{-\sigma}}$ are not equally shifted from the dark state $\ket{\varphi_{1\sigma}}$.
\begin{figure}[h]
\begin{center}
\begin{tabular}{c}
\includegraphics[width=\columnwidth]{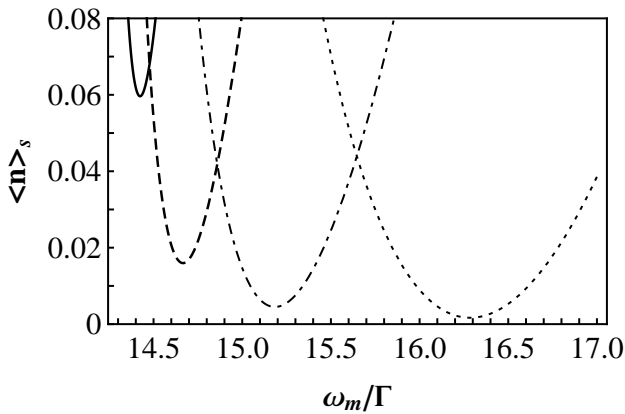}
\end{tabular}
\end{center}
\caption{The steady-state average number of phonons $\langle n\rangle_{s}$ plotted as a function of the mechanical  frequency $\omega_{m}$ for $\alpha=2\Gamma$, an initial temperature $T_{p} =100$ mK (corresponding to $\bar{n}_{p} = 21$), $\gamma_p=2\times10^{-4}\Gamma$, $\Gamma_1=\Gamma_2=\Gamma, T_{1} = T_{2} = 10\Gamma$ with $\Gamma_3 =0.5\Gamma$ and different $\Delta_{2}$: $\Delta_{2}=0.5\Gamma$ (solid line), $\Delta_{2} =\Gamma$ (dashed line), $\Delta_{2} = 2\Gamma$ (dotted-dashed line) and $\Delta_{2} =4\Gamma$ (dotted line).}\label{fig2}
\end{figure}

The above considerations are illustrated in Fig.~\ref{fig2} which shows the steady-state average number of phonons as a function of the frequency $\omega_{m}$ for an initial temperature $T_{p} =100$ mK, corresponding to $\bar{n}_{p} = 21$, equal couplings $T_{1}=T_{2}=T$ and several different values of the detuning $\Delta_{2}$. The minimum of $\langle n\rangle_{s}$ is seen to occur at the frequency $\omega_{m}=\Delta_{2}/2 +\sqrt{2T^{2}+\Delta_{2}^{2}/4}$, which corresponds to the NMR frequency on resonance with the transition from $\ket{\varphi_{1\sigma}}$ to the state $\ket{\varphi_{+\sigma}}$. In the case of large energy difference between dots $3$ and $1$, the mean phonon number $\langle n\rangle_s$ can be reduced significantly. For example, when $\Delta_2=2\Gamma$, the minimum value of the mean phonon number can reach $0.006$, so that we may speak of cooling the NMR to its ground state. Figure~\ref{fig2} also shows that a significant reduction of $\langle n\rangle_s$ is achievable in principle at large $\Delta_{2}$. When $\Delta_{2}=0$, no significant reduction of $\langle n\rangle_{s}$ is observed which makes us to conclude that no cooling can be achieved with three identical quantum dots.
\begin{figure}[h]
\begin{center}
\begin{tabular}{c}
\includegraphics[width=\columnwidth]{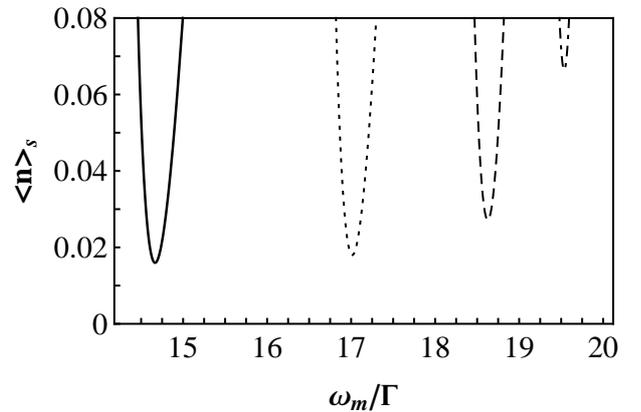}
\end{tabular}
\end{center}
\caption{The mean phonon number $\langle n\rangle_{s}$ as a function the mechanical frequency $\omega_{m}$ for $\alpha=2\Gamma$, an initial temperature $T_{p} =100$ mK (corresponding to $\bar{n}_{p} = 21$), $\Gamma_1=\Gamma_2=\Gamma, \gamma_p = 2\times10^{-4}\Gamma, \Gamma_3 =0.5\Gamma$, $\Delta_{2}=\Gamma$, the average tunnelling amplitude $T=(T_1 +T_2)/2 =10\Gamma$ and different ratios $\beta =T_1/T_2$: $\beta =1$ (solid line), $\beta =4$ (dotted line), $\beta =9$ (dashed line) and $\beta =19$ (dotted-dashed line).}
\label{fig3}
\end{figure}

It is interesting to extend the present analysis to the case of $\tan\eta\neq 1$, corresponding to a dark state with unequal amplitudes of the superposed states $\ket{1_{\sigma}}$ and $\ket{2_{\sigma}}$. A brief look at Fig.~\ref{fig1} and Eq.~(\ref{single}) suggests that a dark state with $\tan\eta > 1$ would be more suitable for cooling since the asymmetry in the population of the dots $1$ and $2$ with $\rho_{22}>\rho_{11}$ could lead to a stronger interaction between the dot $2$ and the NMR. Unequal tunnelling rates with $T_{1}>T_{2}$ is the mechanism for creating~$\rho_{22}>\rho_{11}$. This is shown in Fig.~\ref{fig3}, where we plot the steady-state mean phonon number $\langle n\rangle_s$ as a function of~$\omega_{m}$ for different values of the ratio $\beta = T_{1}/T_{2}$. It is seen that the effect of going from $\beta =1$ to $\beta > 1$ is to reduce rather than to increase the cooling efficiency thus leading to the growing up of the minimum of phonon number. The reason is that unequal tunnelling rates are equally effective in destroying the asymmetry between the cooling and heating frequencies. These considerations imply that the best conditions for cooling with the single-electron dark state are $T_{1}=T_{2}=T$ and $\Delta_{2}\gg \Gamma$.

\subsection{Cooling with the two-electron dark state}

We now consider the cooling mechanism when two electrons can be simultaneously injected into the quantum-dot system. Let us first specialise to the case when only two parallel-spin electrons are present. This situation is described by the Hamiltonian (\ref{hspa}), which in the case of $E_{1}=E_{2}=E$ and $\delta_{1}=\delta_{2}=\delta$ can be written as
\begin{align}
\hat{H}_{\sigma\sigma} = \left(
\begin{array}{ccc}
    0 & T_{1} & - T_{2} \\
    T_{1} & \Delta_{2}\!+\!\delta & 0 \\
    - T_{2} & 0 & \Delta_{2}\!+\!\delta \\
  \end{array}
\right) + (2E\!+\!U_{12})I .\label{tee}
\end{align}
The $3\times 3$ matrix appearing in Eq.~(\ref{tee}) can be diagonalised and the dressed states of the electrons can be identified. However, one can easily verified, without going into the detailed derivation, that the dressed states all would involve the state~$\ket{3_{\sigma}}$, which is dissipative due to its coupling to the drain. This dissipative state is involved regardless of whether the dots are identical $(\Delta_{2}=0)$ or not $(\Delta_{2}\neq 0)$. Thus, no trapping of the electrons could be achieved. Consequently, the electrons would escape to the drain without interacting with the NMR. As a result, no cooling of the NMR could be observed in the interaction with two electrons of the same~spin.

We now generalize the situation to the case of two electrons of opposite spins. To see the advantage of the two-electron over the single-electron case, we assume that dots are identical, i.e. $\Delta_{i}=0 \, (i=1,2,3)$. Notice, that no single-electron cooling was predicted in this regime. Therefore, we may distinguish between one- and two-electron coolings. The case of two electrons of opposite spins is described by the Hamiltonian (\ref{oppo}) which, in general,  has a complicated form but it simplifies in various special cases. We consider two of these cases. A symmetric case in which we choose the charging potential energies $U_{11}=U_{22}=U_{33}$ and $U_{13}=U_{23}=U_{12}$, and an asymmetric case of $U_{11}=U_{22}\neq U_{33}$ and $U_{13}=U_{23}\neq U_{12}$. In the symmetric case the detunings satisfy the condition $\delta_{11}=\delta_{22}=\delta_{33}=\delta_{u}$ and $\delta_{13}=\delta_{23}=0$, whereas in the antisymmetric case $\delta_{11}=\delta_{22}=\delta_{u}$, and $\delta_{13}=\delta_{23}=\delta_{w}\neq 0$. We shall assume additionally that $T_{1}=T_{2}=T$. In each case we will be able to find analytical formulae for the two-electron dressed states of the coupled quantum dots.

\subsubsection{The symmetric case}

The contribution of two opposite-spin electrons to the dynamics of the system is described by the Hamiltonian~(\ref{oppo}), which in the symmetric case simplifies to
\begin{align}
\hat{H}_{\sigma\sigma^{\prime}} =\!\left(
 \begin{array}{ccccccccc}
    \delta_{u} & 0 & 0 & 0 & 0 & T & 0 & T & 0 \\
    0 & 0 & 0 & 0 & 0 & T & 0 & 0 & T \\
    0 & 0 & 0 & 0 & 0 & 0 & T & T & 0 \\
    0 & 0 &  0 & \delta_{u} & 0 & 0 & T & 0 & T  \\
    0 & 0 & 0 & 0 & \delta_{u} & T & T & T & T  \\
    T & T & 0 & 0 & T & 0 & 0 & 0 & 0 \\
    0 & 0 & T & T & T & 0 & 0 & 0 & 0 \\
    T & 0 & T & 0 & T & 0 & 0 & 0 & 0 \\
    0 & T & 0 & T & T & 0 & 0 & 0 & 0 \\
  \end{array}
\right)\!+\!(2E+U_{12})I .\label{sym}
\end{align}

When the detuning $\delta_{u}=0$, a diagonalization of the matrix~(\ref{sym}) leads to dressed states $\ket{{\bold \Phi}}$ which are related to that in the bare-state basis by the transformation
\begin{eqnarray}
\ket{{\bold \Phi}} = \bold{W}_{s}\ket{\bold{\Psi}},
\end{eqnarray}
with the corresponding eigenvalues
\begin{align}
\lambda_1 &=-2\sqrt{2}T,\quad \lambda_2=2\sqrt{2}T,\quad \lambda_3=\lambda_4=-\sqrt{2}T,\nonumber\\
 \lambda_5 &=\lambda_6=\sqrt{2}T,\quad \lambda_7=\lambda_8=\lambda_9 = 0 ,
\end{align}
where $\ket{\bold{\Psi}}$ is a column vector composed of the nine bare states
\begin{align}
|\bold{\Psi}\rangle =& \left[|1_{\uparrow}1_{\downarrow}\rangle,|1_{\uparrow}2_{\downarrow}\rangle,|2_{\uparrow}1_{\downarrow}\rangle,|2_{\uparrow}2_{\downarrow}\rangle,|3_{\uparrow}3_{\downarrow}\rangle,
|1_{\uparrow}3_{\downarrow}\rangle\right. ,\nonumber\\
&\left. |2_{\uparrow}3_{\downarrow}\rangle,|3_{\uparrow}1_{\downarrow}\rangle,|3_{\uparrow}2_{\downarrow}\rangle\right]^{T} ,
\end{align}
and $\bold{W}_{s}$ is a $9\times 9$ transformation matrix whose explicit form is given in the Appendix~\ref{app}.

A careful examination of the transformation matrix ${\bold W}_{s}$, Eq.~(\ref{ws}), reveals that among the dressed states there is a state~$\ket{\Phi_{9}}$ which is a linear superposition involving only bare states of the dots $1$ and $2$:
\begin{align}
\ket{\Phi_9} = \frac{1}{2}\left(|1_{\uparrow}1_{\downarrow}\rangle-|1_{\uparrow}2_{\downarrow}\rangle-|2_{\uparrow}1_{\downarrow}\rangle+|2_{\uparrow}2_{\downarrow}\rangle\right) .\label{dark}
\end{align}
All the remaining states involve bare states of the dot $3$. Since the electrons can tunnel to the drain electrode solely from the dot $3$, this implies that the state $\ket{\Phi_9}$ is an example of a two-electron trapping (dark) state. In other words, two electrons, once trapped in the state $\ket{\Phi_9}$, cannot tunnel into the drain electrode.

We now couple the resulting electron dress states to the mechanical resonator to see if the presence of the two-electron dark state (\ref{dark}) could lead to cooling of the resonator. The interaction between the quantum-dot system and the NMR is determined by the Hamiltonian (\ref{hal}). We make the unitary transformation
\begin{align}
\tilde{H}_{\alpha}(t) = \exp(i\hat{H}_{0}t)\hat{H}_{\alpha}\exp(-i\hat{H}_{0}t) ,
\end{align}
where
\begin{align}
\hat{H}_{0} = \omega_m \hat{a}^{\dagger}\hat{a} +\hat{H}_{d} ,
\end{align}
and find that the Hamiltonian (\ref{hal}) in the dressed-state basis takes the form
\begin{align}
&\tilde{H}_{\alpha}^{(s)}(t) = \frac{\alpha}{2\sqrt{2}}\left\{\hat{a}^{\dagger}\!\left[(|\Phi_3\rangle\!+\!|\Phi_4\rangle)\langle\Phi_9|{\rm e}^{i(\omega_m - \sqrt{2}T)t} +\cdots\right]\right.\nonumber\\
&+\left. \hat{a}\!\left[\left(|\Phi_5\rangle\!+\!|\Phi_6\rangle\right)\!\langle\Phi_9|{\rm e}^{- i(\omega_m - \sqrt{2}T)t} +\cdots\right]\!+ {\rm H.c.}\right\} .\label{epcoupling2}
\end{align}
In writing the expression (\ref{epcoupling2}) we have limited ourselves to those terms that correspond to transitions from the dark state $\ket{\Phi_9}$ and we only concern the case that the frequency of the NMR equals to $\sqrt{2}T$. The remaining terms corresponding to transitions occurring at different frequencies can be neglected for their rapid oscillation. Another kind of terms involving dressed states which contain the state $\ket{3_{\sigma}}$ can be neglected because the populations on these states are negligible.
\begin{figure}[h]
\begin{center}
\includegraphics[width=0.8\columnwidth]{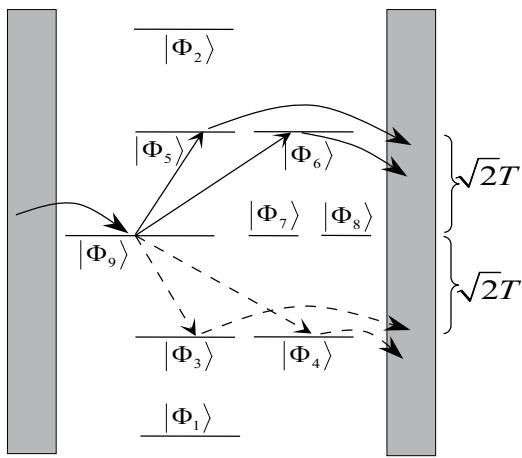}
\end{center}
\caption{Schematic diagram of possible cooling and heating transitions from the dark state $\ket{\Phi_{9}}$ to other dressed states of the system in the case of the symmetric coupling between the dots. The absorption (cooling) processes correspond to the transitions indicated by solid arrows while the emission (heating) transitions are indicated by dashed arrows.}
\label{fig4}
\end{figure}

The Hamiltonian (\ref{epcoupling2}) determines which of the transitions can occur with the absorption of a phonon from the NMR mode and which with the emission of a phonon to the NMR mode.
Figure~\ref{fig4} shows the two-electron dressed states of the quantum dot system with possible transitions from the dark state~$\ket{\Phi_9}$. The electrons trapped in the state $\ket{\Phi_9}$ could make transitions with frequency $\omega_m=\sqrt{2}T$ to the upper states $|\Phi_5\rangle$ and $|\Phi_6\rangle$ by absorbing a phonon from the NMR mode. Of course, the absorption of the phonon would result in a cooling of the NMR to lower temperatures.  Unfortunately, the trapped electrons could also make transitions with the same amplitude and at the same frequency to the lower states $|\Phi_3\rangle$ and $|\Phi_4\rangle$ by emitting a phonon to the NMR mode. Since both processes, the absorption and emission of phonons occur with the same amplitudes and at the same frequencies, no net cooling of the NMR could be achieved. In the following, we will look into the cooling mechanism subject of an asymmetry in the charging potential energies coupling the injected electrons.

\subsubsection{The antisymmetric case}

Our interest is to achieve cooling of the NMR with the help of the two-electron dark state. In order to study this problem, we try to unbalance the symmetry between the absorption and emission processes from the dark state. This could be done, for example, by introducing an asymmetry between the charging potential energies.

Suppose that $U_{11}=U_{22}\neq U_{33}$ and $U_{13}=U_{23}\neq U_{12}$. In this case, $\delta_{13}=\delta_{23}\equiv\delta_{w}\neq 0$. We then find that the Hamiltonian~(\ref{oppo}) reduces to
\begin{align}
\hat{H}_{\sigma\sigma^{\prime}} &=\left(
  \begin{array}{ccccccccc}
    \delta_u & 0 & 0 & 0 & 0 & T & 0 & T & 0 \\
    0 &0 & 0 & 0 & 0 & T & 0 & 0 & T \\
    0 & 0 & 0 & 0 & 0 & 0 & T & T & 0 \\
    0 & 0 &  0 &\delta_u & 0 & 0 & T & 0 & T  \\
    0 & 0 & 0 & 0 &\delta_{33} &T & T & T & T  \\
    T & T &0 & 0 & T&\delta_{w}& 0 & 0 & 0 \\
    0 & 0 & T & T & T&0 &\delta_{w}& 0 & 0 \\
    T & 0 & T & 0 & T&0 & 0 &\delta_{w}& 0 \\
    0 & T & 0 & T & T&0 & 0 & 0 &\delta_{w} \\
  \end{array}
\right) \nonumber\\
 &+ (2E + U_{12})I .\label{ant}
\end{align}

A diagonalization of the Hamiltonian (\ref{ant}) produces complicated formulae for the eigenvalues (energies) and the eigenvectors (dressed states) of the system. Even at $\delta_{u}=0$ there are still three parameters involved in these formulae,~$T, \delta_{33}$ and $\delta_{w}$. Therefore, for a simplification of the formulae, we express $\delta_{33}$ and $\delta_{w}$ in terms of~$T$. For example, setting the detunings $\delta_{33}=\delta_{u}+2T$, $\delta_w=T$ and $\delta_{u}=0$, we find simple expressions for the eigenvalues
\begin{align}
\lambda_1 &= 4T,\quad \lambda_2 = \lambda_3 = 2T, \quad \lambda_4 = \lambda_5 =T,\nonumber\\
\lambda_6 &= -2T,\quad \lambda_7=0,\quad \lambda_8 =\lambda_9 = -T ,
\end{align}
and the corresponding dressed states
\begin{eqnarray}
\ket{{\bold \Phi}} = \bold{W}_{a}\ket{\bold{\Psi}} ,\label{equ37}
\end{eqnarray}
where $\bold{W}_{a}$ is a $9\times 9$ transformation matrix whose explicit form is given in the Appendix~\ref{app}.

An inspection of Eq.~(\ref{wa}) reveals that one of the dressed states,~$\ket{\Phi_{7}}$, is of the same form as the state (\ref{dark}), i.e. the state~$\ket{\Phi_{7}}$ is a dark state involving bare states of only the dots~$1$ and $2$.  We have checked that independent of the choice of~$\delta_{33}$ and $\delta_{w}$, there is always an asymmetric dressed state that involves bare states of the dots~$1$ and~$2$ only. Hence $\delta_{u}=0$ is the general condition for the creation of the dark state in the system.

Given the electron dressed states, we may transform the interaction Hamiltonian $\hat{H}_{\alpha}$ into the dressed-state basis and find
\begin{align}
&\tilde{H}_{\alpha}^{(a)}(t) = \frac{\alpha}{\sqrt{6}}\left\{\hat{a}^{\dagger}\!\left[(|\Phi_9\rangle\!+\!|\Phi_8\rangle)\langle\Phi_7|{\rm e}^{i(\omega_m - T)t}+\cdots\right]\right. \nonumber\\
&\left. - \frac{\hat{a}}{\sqrt{2}}\!\left[(|\Phi_2\rangle\!+\!|\Phi_3\rangle)\langle\Phi_7|{\rm e}^{-i(\omega_m - 2T)t}\!+\!\cdots\right]\!+ {\rm H.c.}\!\right\} ,\label{epcoupling3}
\end{align}
where as before for the symmetric case, we have explicitly listed only the dominant terms in the coupling of the dressed system to the NMR. One can notice that these dominant terms involve operators describing transitions to and from the dark state $\ket{\Phi_{7}}$.
\begin{figure}[h]
\begin{center}
\begin{tabular}{c}
\includegraphics[width=0.8\columnwidth]{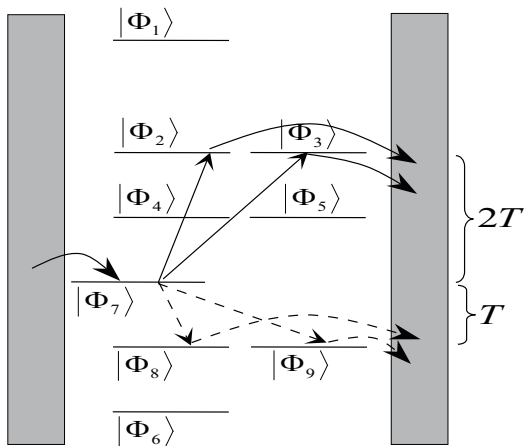}
\end{tabular}
\end{center}
\caption{Two-electron dressed states of the quantum-dot system in the case of the asymmetric coupling between the dots. The solid arrows indicate dominant cooling transitions from the dark state $|\Phi_7\rangle$ to the upper states $|\Phi_2\rangle$ and $|\Phi_3\rangle$, and dashed arrows indicate emission (heating) transitions to the lower energy states $|\Phi_8\rangle$ and $|\Phi_9\rangle$.}
\label{fig5}
\end{figure}

Figure~\ref{fig5} shows the two-electron dressed states of the quantum-dot system for an asymmetric coupling between the dots. The arrows indicate transitions from the dark state that are induced by the coupling of the dressed system to the NMR. We see that the dark state $|\Phi_7\rangle$ is strongly coupled to only four states of the dressed system, two upper and two lower energy states. Thus, similar to the symmetric case both cooling and heating processes can occur. However, a notable distinction between the symmetric and antisymmetric cases is that in the asymmetric case the absorption and emission processes occur at different frequencies. It is easily verified from Eq.~(\ref{epcoupling3}) that the absorption processes occur at frequency $\omega_{m}=2T$, whereas emission processes occur at $\omega_{m}=T$. This means that at the frequency $\omega_{m}=2T$, the absorption of a phonon from the NMR is not accompanied by an emission of a phonon of the same frequency. As a consequence the NMR can be cooled to its ground state.

\begin{figure}[h]
\begin{center}
\includegraphics[width=\columnwidth]{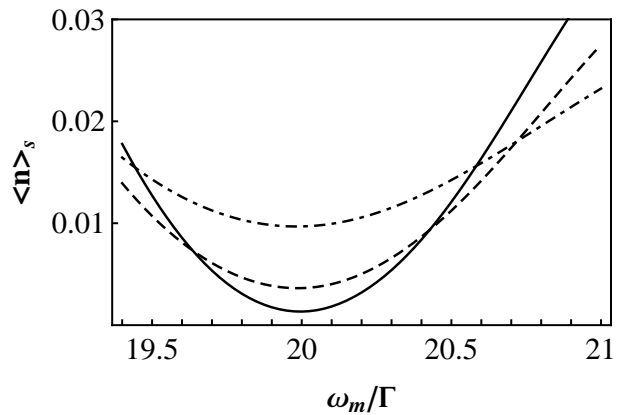}
\end{center}
\caption{Dependence of the steady-state mean phonon number $\langle n\rangle_{s}$ on frequency $\omega_{m}$ for the asymmetric case of $\delta_{w}=T$ with an initial temperature $T_{p} =100$ mK (corresponding to $\bar{n}_{p} = 21$), $\Gamma_1=\Gamma_2=\Gamma$, $T=10\Gamma, \alpha=2\Gamma$, $\delta_u=0$, $\Delta_i=0 \,(i=1,2,3), \gamma_p=2\times10^{-4}\Gamma, \delta_{33}=2T$, and different tunnelling rates $\Gamma_3$: $\Gamma_{3}=0.5\Gamma$ (solid line), $\Gamma_3 =\Gamma$ (dashed line), $\Gamma_{3}= 2\Gamma$ (dotted-dashed line).}
\label{fig6}
\end{figure}

This is illustrated in Fig.~\ref{fig6} which shows the steady-state mean phonon number $\langle n\rangle_{s}$ as a function of the frequency~$\omega_{m}$ of the NMR for different tunnelling rates of the electrons to the drain. Cooling of the resonator shows up as a reduction of the mean phonon number with a clear minimum at $\omega_{m}=2T$. The minimum of $\langle n\rangle_{s}$, corresponding to the maximum of the cooling of the NMR, is achieved at small values of the tunnelling rate $\Gamma_3$ and the effect of increasing $\Gamma_{3}$ is seen to reduce the cooling efficiency. Taking into account the mean phonon number of $\bar{n}_p=21$ in the initial equilibrium thermal state, the value of $\langle n\rangle_s\simeq1.3\times10^{-3}$ achieved with the two electron dark state means that the NMR has been cooled to its ground state. With the practical values of the frequency $\omega_{m}=100$ MHz, the predicted minimum value of  $\langle n\rangle_s\simeq1.3\times10^{-3}$ corresponds to a temperature of the NMR reduced to $T_{m}\simeq0.72$ mK.

\begin{figure}[htp]
\begin{center}
\includegraphics[width=\columnwidth]{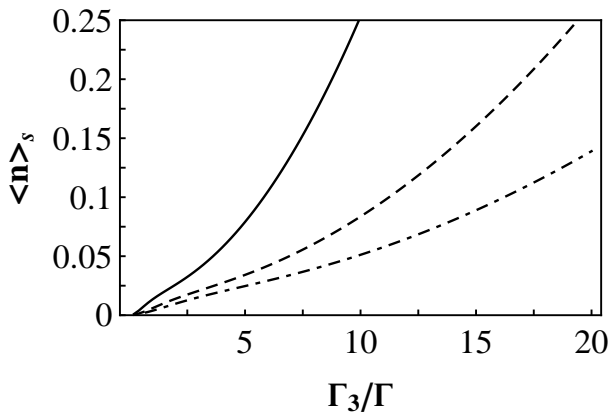}
\end{center}
\caption{The dependence of minimum value of $\langle n\rangle_{s}$ on the tunnelling rate $\Gamma_3$ for the asymmetric case of $\delta_{w}=T$ with an initial temperature $T_{p} =100$ mK (corresponding to $\bar{n}_{p} = 21$), $\Gamma_1=\Gamma_2=\Gamma$, $\omega_m=2T$, $\alpha = 2\Gamma$, $\delta_u=0$, $\Delta_i=0 \, (i=1,2,3)$, $\gamma_p = 2\times10^{-4}\Gamma$, $\delta_{33}=2T$, and different $T$: $T=5\Gamma$ (solid line), $T=10\Gamma$ (dashed-line), $T=15\Gamma$ (dashed-dotted line).}
\label{fig7}
\end{figure}

Figure~\ref{fig7} shows the effect of the tunnelling rate $\Gamma_{3}$, at which electrons escape from the system, on the magnitude of the minimum value of $\langle n\rangle_{s}$. The effect of increasing $\Gamma_{3}$ is clearly to reduce the minimum value of the mean number of phonons. For small tunnelling rate $T$ the increase of the minimum value of $\langle n\rangle_{s}$ with $\Gamma_{3}$ is considerably more rapid than it is for large~$T$. This is readily understood if one recalls that increasing of the escape rate of the electrons results in a shortening of the interaction time of the electrons with the NMR. The shorter interaction time implies less of the energy taken from the~NMR. Thus, the minimum value of $\langle n\rangle_{s}$ degradates with the increasing tunnelling rate $\Gamma_3$.

The above analysis of the cooling mechanism were based on the energy structure of the system of the coupled dots. Further insight into the role of the two electron dark state in the cooling process is gained by considering the dynamics of the dressed states $\ket{\Phi_{i}}$. Because it is precisely the effect of trapping of the population in the state $\ket{\Phi_{7}}$ that was crucial for the cooling mechanism, we derive the equation of motion for the population of the state $\ket{\Phi_{7}}$ that is represented by the density matrix element~$\rho_{77}$. To illustrate that the dressed states that involve states of the dot $3$ cannot be trapping states, we also derive equations of motion for the dressed states $\ket{\Phi_{2}}$ and $\ket{\Phi_{3}}$ that according to Eq.~(\ref{equ37}) are linear superpositions involving the state $\ket{3_{\sigma}}$.

Consider the evolution of the quantum dot system alone, i.e. in the absence of the NMR. The evolution is determined by the master equation that is obtained from Eq.~(\ref{master}) by tracing over the NMR states
\begin{eqnarray}
\frac{d}{dt}\rho = -i[\hat{H}_{d},\rho]+\mathcal {L}_e\rho ,\label{mqd}
\end{eqnarray}
where $\hat{H}_{d}$ is given in Eq.~(\ref{Hd2}) and $\mathcal {L}_e\rho$ is given in Eq.~(\ref{equ13}).

When we project the master equation (\ref{mqd}) onto $\ket{\Phi_{i}}\, (i=2,3,7)$ on the right and $\bra{\Phi_{i}}$ on the left, we obtain the following equations of motion
\begin{align}
\dot{\rho}_{77} &= \Gamma\left(\rho_{\varphi_{1\downarrow 1\downarrow}} +\rho_{\varphi_{1\uparrow 1\uparrow}}\right) -\frac{1}{3}i\delta_{u}\left(\tilde{\rho}_{17} +2\tilde{\rho}_{57} +2\tilde{\rho}_{67}\right)  ,\nonumber\\
\dot{\rho}_{22} &= -\frac{2}{3}\Gamma_{3}\rho_{22} +\frac{1}{6}\Gamma \left[2\rho_{\varphi_{1\downarrow 1\downarrow}} +\left(1+\sqrt{2}\right)^{2}\rho_{\varphi_{+\uparrow +\uparrow}}\right. \nonumber\\
&\left. +\left(\sqrt{2}-1\right)^{2}\rho_{\varphi_{-\uparrow -\uparrow}}\right]
-\frac{1}{3}i\delta_{u}\!\left[\tilde{\rho}_{23} +\sqrt{2}\!\left(\tilde{\rho}_{28} +\tilde{\rho}_{29}\right)\right]  ,\nonumber\\
\dot{\rho}_{33} &= -\frac{2}{3}\Gamma_{3}\rho_{33} +\frac{1}{6}\Gamma \left[2\rho_{\varphi_{1\uparrow 1\uparrow}} +\left(1+\sqrt{2}\right)^{2}\rho_{\varphi_{+\downarrow +\downarrow}}\right. \nonumber\\
&\left. +\left(\sqrt{2}-1\right)^{2}\rho_{\varphi_{-\downarrow -\downarrow}}\right]
-\frac{1}{3}i\delta_{u}\!\left[\tilde{\rho}_{23} +\sqrt{2}\!\left(\tilde{\rho}_{38} +\tilde{\rho}_{39}\right)\right]  ,\label{equ39}
\end{align}
where $\rho_{\varphi_{1\sigma 1\sigma}}$ is the population of the single electron state~$\ket{\varphi_{1\sigma}}$ and $\tilde{\rho}_{ij} ={\rm Im}\rho_{ij}$ is the imaginary part of the coherence between states $i$ and $j$.

We see from Eq.~(\ref{equ39}) that the dressed state $\ket{\Phi_{7}}$ does not decay, but can be populated by transitions from the single electron states $\ket{\varphi_{1\uparrow}}$ and $\ket{\varphi_{1\downarrow}}$. In contrast, the states $\ket{\Phi_{2}}$ and $\ket{\Phi_{3}}$ that involve the state of the dot $3$ decay with the rate $2\Gamma_{3}/3$. The population state $\rho_{77}$ is coupled to the other state through the detuning $\delta_{u}$.  In the steady state $(\dot{\rho}_{ii}=0)$ with $\delta_{u}=0$, the population can be completely transferred to the state $\ket{\Phi_{7}}$. Thus, the equations of motion (\ref{equ39}) clearly demonstrate that in the steady-state with $\delta_{u}=0$, the state~$\ket{\Phi_{7}}$ becomes a trapping state.

The expressions (\ref{equ39}) also show that the single electron states play the major role in the two electron cooling process. The dark state $\ket{\Phi_{7}}$ is populated with the rate $\Gamma$ from the single electron states $\ket{\varphi_{1\uparrow}}$ and $\ket{\varphi_{1\downarrow}}$. When $\delta_{u}=0$, no population can be transferred from the other two-electron states.

We conclude this section by pointing out that the conditions predicted by Fig.~\ref{fig6} for cooling of a NMR with the two-electron dark state may be met in the current experimental situations. The value of the tunnelling rate $\Gamma/2\pi =5$\, MHz used to generate Fig.~\ref{fig6} corresponds to the typical values occurring in electron transport experiments where the tunnelling rates ranging from 10 kHz to 10 GHz were measured~\cite{Le99,Rev}. The predicted cooling of the NMR occurs at frequencies $\omega_{m}/\Gamma \approx 20$ corresponding to the value of $\omega_{m}\approx 100$ MHz, which appears to be practical~\cite{APL92}. Note further that the expected cooling times $t_{c}$ correspond to that required for the mean phonon number to reach the steady-state, which according to Eq.~(\ref{dn}) are of the order of $(\gamma_{p}+A_{-}-A_{+})^{-1}$. With the parameter values of $\gamma_{p}=2\times10^{-4}\Gamma$, corresponding to a realistic quality factor $Q\sim10^{5}$~\cite{qual1,qual2}, $A_{-}\simeq4.577\Gamma$ and $A_{+}\simeq5\times10^{-3}\Gamma$, the cooling could be achieved at times $t_{c}\simeq4.4\times10^{-8}$\, s.

Finally, interesting further generalizations of the results presented in this paper could include other configurations of the quantum dot system such us $V$-type or linear systems. It is also worth considering systems composed of a large number of regularly distributed quantum dots.

\section{Summary}\label{sc4}

We have studied the dark state cooling technique in a system composed of three coupled quantum dots interacting with an NMR. Two cases of coupling of the NMR to single and two-electron dark states have been considered. Cooling of the NMR to its ground state can occur in both cases. In the case of the interaction of the resonator with a single-electron dark state, no cooling of the resonator occurs unless the quantum dots are not identical. The steady-state cooling is possible only if the energy state of the quantum dot coupled to the drain electrode is detuned from the energy states of the dots coupled to the electron source electrode. In the case of identical dots the cooling and heating processes were found to occur with the same amplitudes and at the same frequency. This results in no effective cooling of the NMR. When the dots are not identical, the phonon emission (heating) and phonon absorption (cooling) transitions  occur at different frequencies giving rise to an effective cooling of the NMR.

In the case of the interaction of the NMR with the two-electron dark state, we have showed that an effective cooling can be achieved even with identical quantum dots subject of asymmetries only in the charging potential energies coupling the injected electrons. In the symmetric case, we have shown that the cooling and heating transitions occur with the same amplitudes and at the same frequencies. An asymmetry in the charging potential energies is found to lead to unequal shifts of the effective dressed states of the systems which results in the cooling and heating frequencies to become detuned from each other. This shows that  the physics of the  two-electron dark state cooling process is quite similar to that of the single-electron dark state cooling. However, there is an important difference that these two cooling processes occur at significantly different frequencies. This indicates that the frequency at which the resonator could be cooled to its ground state can be changed by switching from the one-electron to the two-electron Coulomb blockade process.

\section*{Acknowledgments}
This work is supported by the National Basic Research Program of China (Grant No. 2012CB921602), the National Natural Science Foundation of China (Grant No. 11074087), the Natural Science Foundation of Hubei Province (Grant No. 2010CDA075), and the Nature Science Foundation of Wuhan City (Grant No. 201150530149).

\appendix
\section{Transformation matrices}~\label{app}
In this Appendix we give the transformation matrices from the bare electron states to the two-electron dressed states. In the case of the symmetric coupling the transformation matrix is of the form
\begin{widetext}
\begin{align}
&\bold{W}_s=\frac{1}{4}\left(
  \begin{array}{ccccccccc}
    -1& -1 & -1 & -1 & -2 & \sqrt{2} & \sqrt{2} & \sqrt{2} & \sqrt{2} \\
    1 &1& 1 & 1 & 2 &\sqrt{2} & \sqrt{2} & \sqrt{2} & \sqrt{2} \\
    \sqrt{2} & -\sqrt{2} & \sqrt{2} & -\sqrt{2} & 0 &0& 0 & -2 & 2 \\
    \sqrt{2} & \sqrt{2}& -\sqrt{2} & -\sqrt{2} & 0&-2& 2 & 0 & 0 \\
    -\sqrt{2} & \sqrt{2} & -\sqrt{2} & \sqrt{2} & 0&0& 0 & -2 & 2  \\
    -\sqrt{2} & -\sqrt{2} & \sqrt{2} & \sqrt{2} & 0&2& 2 & 0 & 0  \\
    0 & 0 & -2 & 0 & 0& -2 &0 &2& 2  \\
    -\sqrt{2} & -\sqrt{2} & -\sqrt{2} & -\sqrt{2} & 2\sqrt{2}&0& 0 & 0 & 0 \\
    -2 & 2 & 2 & -2 &0& 0 &0 & 0 & 0 \\
  \end{array}
\right) ,\label{ws}
\end{align}
whereas in the case of the asymmetric coupling, the matrix is given by
\begin{align}
&\bold{W}_a=\frac{1}{6}\left(
  \begin{array}{ccccccccc}
    1& 1 & 1 & 1 & 4 &2 & 2 & 2 & 2 \\
    -\sqrt{3} &\sqrt{3}& -\sqrt{3} & \sqrt{3} & 0 & 0& 0 & -2\sqrt{3} & 2\sqrt{3} \\
    -\sqrt{3} & -\sqrt{3} & \sqrt{3} & \sqrt{3} & 0&-2\sqrt{3}& 2\sqrt{3} & 0 & 0 \\
    0 & 0 & 0 & 0 & 0&-3& -3 & 3 & 3  \\
    -2 & -2 & -2 & -2 & 4 &-1& -1 & -1 & -1 \\
    -2 & -2 & -2 & -2 & -2&2& 2 & 2 & 2 \\
    -3 & 3 & 3 & -3 &0&0 & 0 & 0 & 0 \\
    \sqrt{6} & -\sqrt{6} &\sqrt{6} & -\sqrt{6} & 0&0& 0 & -\sqrt{6} & \sqrt{6} \\
    \sqrt{6} & \sqrt{6} & -\sqrt{6} & -\sqrt{6} &0&-\sqrt{6} & \sqrt{6} & 0 & 0  \\
  \end{array}
\right) .\label{wa}
\end{align}
\end{widetext}

\end{document}